\documentstyle[12pt,fleqn]{article}
\advance\textheight by \topskip
\newcommand{\tma}[2]{\begin{#1} \label{se:#2}}
\newcommand{\tme}[2]{\end{#1}}
\newcommand{\rsec}[2]{\section{#1} \label{se:#2}}
\newcommand{\Proo}{{\bf Proof}\\}
\newcommand{\Prooe}{\phantom{.}\hfill $\Box$}
\newcommand{\rse}[1]{\ref{se:#1}}

\newcommand{\ren}[1]{\ref{en:#1}}
\newcommand{\ds}{\displaystyle}
\newcommand{\Rset}{{\rm I}\!{\rm R}}
\newcommand{\Eset}{{\rm I}\!{\rm E}}
\newcommand{\Nset}{{\rm I}\!{\rm N}}

\newcommand{\htP}{\hat{P}}
\newcommand{\htPp}{\hat{p}}
\newcommand{\htQ}{\hat{Q}}
\newcommand{\htS}{\hat{S}}
\newcommand{\htSs}{\hat{s}}
\newcommand{\hta}{\hat{+}}
\newcommand{\htm}{\hat{\cdot}}
\newcommand{\htN}{\hat{0}}
\newcommand{\htE}{\hat{1}}
\newcommand{\htdel}{\hat{\delta}}
\newcommand{\htSu}{\widehat{\sum}}
\newcommand{\htPr}{\widehat{\prod}}
\newcommand{\htIn}{\widehat{\int}}
\newcommand{\htRt}[2]{\widehat{\sqrt[#1]{#2}}}

\begin{document}

\begin{center}
 \LARGE\bf Uncertain dynamical systems defined by pseudomeasures\\
 \vspace{2cm}
 \Large\sf Andreas Hamm\\ \large Fachbereich Physik\\Universit\"{a}t GH
                                    Essen\\ 
           45117 Essen\\Germany\\
  \vspace{0.5cm}
 \small\rm July 1996, revised in February 1997
\end{center}

\[ \]
\vspace{-1cm}

\thispagestyle{empty}

\normalsize

\begin{abstract}
This paper deals with uncertain dynamical systems
in which predictions about the future state of a system
are assessed by so called pseudomeasures. Two special cases
are stochastic dynamical systems, where the pseudomeasure is
the conventional probability measure, and fuzzy dynamical systems in which
the pseudomeasure is a so called possibility measure.

New results about possibilistic systems and their relation to 
deterministic and to stochastic systems are derived by using
idempotent pseudolinear algebra.

By expressing large deviation estimates for stochastic perturbations
in terms of possibility measures, we obtain a new interpretation of the
Freidlin-Wentzell quasipotentials for sto\-chastic perturbations of dynamical
systems as invariant possibility densities.

\end{abstract}

\newpage

\rsec{Introduction}{1}

Modelling natural processes by deterministic dynamical systems requires
usually simplifying approximations and assumptions. It is reasonable to
look for methods which take into account the uncertainties caused by
these inevitable simplifications.

The historically oldest method is probability theory: If the uncertainties can
be interpreted as the cumulative effect of a large number of
independent small perturbations the rules of probability theory can be
applied to estimate how often in a large sample of identical processes
an actually occuring event would be close to the deterministically
predicted event within certain bounds. The arguably most successful
results of this strategy can be found in statistical mechanics.

About 30 years ago, L.~Zadeh \cite{Za1} introduced a different
approach to uncertainties: 
The theory of fuzzy sets. This theory has become increasingly popular as a
successful tool for modelling uncertainties in various applications, notably
in engineering (process control) and information technology (expert systems).

The advantages of this approach when compared to probability theory are a
higher flexibility of rules, an intuitive appeal, and some computational 
merits. These advantages are favourable for the handling of uncertainties in
single events for which no statistical information is available, and for
quantifying semantic statements about uncertainties. On the other hand there
are bitter controversies about the epistemological justification of fuzzy
methods. But although in some applications of fuzzy tools a hint of
arbitrariness can still be detected, there are now well developed systematic
ways of using the fuzzy approach, in particular the branch called
possibility theory \cite{DP1}.

It is not the intention of this article to explain or to justify the fuzzy
approach, and there can be little doubt that it will never reach the same
importance in theoretical physics as the probabilistic approach. But the
existence of non-probabilistic concepts of uncertainty suggests that some
questions that are asked about the effect of random perturbations on dynamical
systems can be posed in a wider framework. Such questions belong often to one
of the following classes:
\begin{itemize}
 \item Stability questions: Which of the features of a deterministic dynamical
       system are robust to small uncertainties?
 \item Asymptotics of weak perturbations: Are there approximations for the
       deviations of an uncertain system from a deterministic system when the
       uncertainties are small?          
\end{itemize}

It is obvious that it makes sense to try to answer questions of the first type
in a way which is as independent of a special model of uncertainty as
possible. Here, the results of Section \rse{5} about the sets of states which
dominate the long-term behaviour in an uncertain system are an interesting
example. It will turn out that these sets
are roughly identical to the chain recurrent sets \cite{Ru1} of the underlying
deterministic system --- independent of whether the uncertainties are modelled
according to probability theory or according to possibility theory.

On the other hand, questions of the second type do not ask for such generality
and model independence, as they usually refer to concrete situations. There
seems to come little direct motivation from concrete physical problems to
study possibilistic systems. However, we will show in Section
\rse{6} that possibilistic systems can be used as a tool for finding
approximate results
about probabilistic systems and for solving variational problems. This insight
forms a new motivation for
theoretical physicists to look at possibilistic methods, and it bridges the
gulf between probability theory and possibility theory. We mention already
here that the most interesting link between probability theory and
possibility theory is established by the much used estimates of large
deviation type \cite{Va1,DZ1}.
In the context of dynamical systems this means that the so called
quasipotentials or nonequilibrium potentials \cite{FW1,Gr1,Ki1,HG1}
--- a standard tool for studying stochastic perturbations  ---
have a natural interpretation in the context of possibility theory.
 
From the start we will introduce the non-probabilistic approaches to
uncertainty not as a contrast but as a generalisation of the probabilistic
approach. This is made easy by following not Zadeh's original way of
modifying set theory (or logic), but by
generalising the notions of measure and integral. Several authors have
suggested such 
generalisations (e.g.\cite{We1,SM1,De1}); we follow most closely
Sugeno's work \cite{SM1} on ``pseudo-additive 
measures''. Our definitions differ slightly from Sugeno's in that we stress
the algebraic properties of the ``pseudo-additive integrals''. Since we
introduce several generalisations of well-known mathematical objects we find
it convenient to use the prefix ``pseudo'' to name the generalised versions.

The algebraic properties of pseudointegrals bring us to another essential
ingredient of this paper: Pseudolinear algebra. Over the last years there have
been increasingly many applications of this interesting branch of
mathematics (e.g. in finite automata \cite{Ga1}, morphology neural
networks \cite{DH1}, image processing \cite{Ri1}, and one dimensional
crystallographic models \cite{CG1}), indicating that it is worth while to
advocate a wider spread of its ideas. A particularly well studied
special case of pseudolinear algebra is the so called idempotent
algebra which has its roots in optimisation theory. There seem to be
several authors which have discovered its main results independently
(for reviews see \cite{Cu1,GM1,Zi1,MS1}).  

While most of the existing results in idempotent pseudolinear algebra refer to
finite dimensional situations, we need infinite dimensional versions. Section
\rse{4} contains our results on the pseudolinear eigenvalue problem on a
function space. 
Recently some similar and some stronger results became available through
\cite{MS1}, and some of the results have been known in the language of infinite
horizon problems in optimisation theory for a while \cite{Le1}. Nevertheless, 
a complete presentation of our approach is important here,
not only for the sake of self-consistency of this paper but because our
approach --- unlike \cite{MS1} --- concentrates on the structure of the
eigenfunctions which will
have the meaning of invariant possibility densities or quasipotentials in our
applications. 

The material in this paper is organised as follows:

Section \rse{2} introduces the definitions of semirings, pseudomeasures, and
pseudointegrals. In Section \rse{3} we explain how these concepts can be used
to model uncertain dynamical systems and show that invariant pseudodensities
for the characterisation of their long-term behaviour fulfill a pseudolinear
eigenvalue equation. Section \rse{4} contains the solution of this eigenvalue
problem under the algebraic assumption of idempotency of pseudoaddition. This
assumption is fulfilled if the pseudomeasures are possibility measures.
Section \rse{5} is devoted to the study of links between possibilistic
and deterministic systems, and we will show a connection between solutions of
the pseudolinear eigenvalue problem and chain recurrent sets of deterministic
systems. Finally, in Section \rse{6} we will use possibilistic systems to
derive results about stochastic systems. In particular we will formulate the
so called large deviation property in terms of possibility measures and
explain the connection to quasipotentials.

\rsec{Pseudomeasures and pseudointegrals \protect\\over semirings}{2}

Let $(X,{\cal A})$ be a measurable space (i.e. ${\cal A}$ is a
$\sigma$-algebra of subsets of $X$). A pseudomeasure is a certain set
function $\htP:{\cal A}\rightarrow E$ which is introduced in order to
assess the guess that an element of $X$ belongs to $A\in{\cal A}$ at
$\htP(A)$. We give some structure to the evaluation set $E$ which makes
it suitable for such assessments.

\tma{Def}{2-1}
An ordered commutative monoid is a quadruple $(E,\leq,\hta,\htN)$
with
\begin{enumerate}
\item      $(E,\leq)$ is a partially ordered set (i.e. there is a
           reflexive, transitive, and antisymmetric relation $\leq$ on
           $E$).
\item         $(E,\hta)$ is a commutative semigroup (i.e. $\hta$ is an
              associative and commutative binary operation on $E$). The
              operation $\hta$ is called pseudoaddition.
\item          $\htN$ is an identity element of $E$ under $\hta$ (i.e.
               $a\hta \htN = a$ for all $a\in E$).
\item         $\hta$ is a monotone operation on $(E,\leq)$ (i.e.
              $a\leq b$ for $a,b\in E$ implies $a\hta c \leq b\hta c$
              for all $c\in E$).
\end{enumerate}
$(E,\leq,\hta,\htN)$ is called positively ordered commutative monoid if
in addition $\htN \leq a$ for all $a\in E$.
\tme{Def}{2-1}

For questions of convergence it is most convenient to deal with a
special type of partially ordered sets:

\tma{Def}{2-2}
A partially ordered set $(E,\leq)$ in which each subset $G\subset E$ has
a least upper bound $\sup G$ as well as a greatest lower bound $\inf G$
is called a complete lattice.

A partially ordered set $(E,\leq)$ in which for each pair $(a,b)\in
E\times E$ either $a\leq b$ or $b\leq a$ (i.e., $a$ and $b$ are
comparable) is called totally ordered.
\tme{Def}{2-2}

\tma{Exa}{2-3}
\rm
Simple but important examples of complete lattices are
\begin{enumerate}
 \item   $I=[0,1]$,
 \item   $\Eset=\Rset\cup \{-\infty,\infty\}$,
 \item   $\Eset^{+}=\{ a\in \Rset: a\geq 0 \} \cup \{\infty\}$,
 \item   $\Eset^{-}=\{ a\in \Rset: a\leq 0 \} \cup \{-\infty\}$
\end{enumerate}
with the usual order relation $\leq$ for real numbers. These examples
are all totally ordered.

For every complete lattice $(E,\leq)$, $(E,\leq,\hta,\htN)$ with
$\htN=\inf E$ and $a\hta b = \sup \{a,b\}$ is a positively ordered
commutative monoid.

Two other examples for pseudoadditions are $a\hta b = a+b$ on
$E=\Eset^+$ and $a\hta b = a+b-ab$ on $E=I$.

Pseudoadditions in positively ordered commutative monoids, especially
for $E=I$, are called triangular conorms and have been studied in
great detail in the context of probabilistic metric spaces \cite{SS1}.
\tme{Exa}{2-3}

\tma{Def}{2-4}
The limit superior of a sequence $(a_i)_{i=1,2,\dots}$ of elements of
a complete lattice is defined as
\[
 \overline{\lim}\, a_i = \inf_j \sup_{i\geq j} a_i ,
\]
the limit inferior as
\[
 \underline{\lim}\, a_i = \sup_j \inf_{i\geq j} a_i .
\]
If $\overline{\lim}\, a_i = \underline{\lim}\, a_i = a_*$ then $(a_i)$
is called order convergent to the limit $\ds a_*=\lim_i a_i$.
\tme{Def}{2-4}

From now on we assume that the pseudoaddition is continuous:
\[
   \lim_i (a_i \hta b_i) = \lim_i\ a_i  \hta  \lim_i\ b_i
\]

\tma{Def}{2-5}
Let $(E,\leq)$ be a complete lattice, $(E,\leq,\hta,\htN)$ a positively
ordered commutative monoid, and $(X,{\cal A})$ a measurable space.
A pseudomeasure is a set function $\htP:{\cal A}\rightarrow E$
with the properties
\begin{enumerate}
\item      $\htP(\emptyset) = \htN$,
\item       $\htP \left( \bigcup_i A_i \right) =
             \htSu_i \htP (A_i) $ \ for every family $(A_i)$ of
            pairwise disjoint $A_i \in {\cal A}$.
\end{enumerate}
Here, the obvious symbol $\htSu$ for pseudosums was used.

Sets of pseudomeasure $\htN$ are called $\htP$-nullsets. Statements
which are true for all $x\in X-O$ with $O$ a $\htP$-nullset are said to
hold ($\htP$-)almost everywhere.
\tme{Def}{2-5}

Having defined a generalisation of measures, we next generalise integrals.
Again we are led by the aim to retain some of the algebraic properties of the
conventional integral. Therefore we introduce a further operation on $E$.

\tma{Def}{2-6}
A positively ordered commutative semiring is the collection
\linebreak
$(E,\leq,\hta,\htN,\htm,\htE)$ where $\htN \neq \htE$ and
\begin{enumerate}
\item      $(E,\leq,\hta,\htN)$ is a positively ordered commutative
           monoid.
\item       $(E,\leq,\htm,\htE)$ is an ordered commutative monoid.
            The operation $\htm$ is called pseudomultiplication.
\item          $(a\hta b)\htm c = a\htm c \hta b\htm c$ for all
               $a,b,c\in E$.
\item         $\htN \htm a = \htN$ for all $a\in E$.
\end{enumerate}
If in addition $a\htm b=\htN \Rightarrow a=\htN$ or $b=\htN$ then
the semiring is called entire.

If $a\htm c = b\htm c  \Rightarrow  a = b$ for all $a,b,c\in E$, $c\neq
\htN$ then the semiring is called cancellative.
\tme{Def}{2-6}

In the following we assume continuity of the pseudomultiplication:
\[
   \lim_i (a_i \htm b_i) = \lim_i\ a_i  \htm  \lim_i\ b_i  .
\]
From now on, we deal only with entire semirings.

\tma{Exa}{2-7}
\rm
For every complete lattice $E$, $(E,\leq,\hta,\htN,\htm,\htE)$ with
$\htN=\inf E$, $\htE=\sup E$, $a\hta b = \sup \{a,b\}$, and $a\htm b =
\inf \{a,b\}$ is an ordered commutative semiring.

Some examples for semirings involving the lattices of Example \rse{2-3}
are:
\begin{enumerate}
 \item   $E=\Eset^+, \htN = 0, \htE = 1,
          a \hta b = a + b, a \htm b = ab$         \label{en:2-7.1}
 \item   $E=I, \htN = 0, \htE = 1,
          a \hta b = \max\{a,b\}, a \htm b = ab$   \label{en:2-7.2}
 \item   $E=I, \htN = 0, \htE = 1,
          a \hta b = \max\{a,b\}, a \htm b = \min\{a,b\}$
          \label{en:2-7.3}
 \item   $E=\Eset^-, \htN = -\infty, \htE = 0,
          a \hta b = \max\{a,b\}, a \htm b = a + b$  \label{en:2-7.4}
\end{enumerate}

There is a close relation between Examples \ren{2-7.2} and
\ren{2-7.4}: Because of \linebreak $\exp (\max\{a,b\}) =
\max\{\exp(a),\exp(b)\}$, 
$\exp(a+b)=\exp(a) \exp(b)$, $\exp (-\infty) = 0$, and $\exp (0) = 1$,
the exponential function is a semiring morphism between those examples.

All the Examples \ren{2-7.1} to \ren{2-7.4} are entire semirings.
Examples \ren{2-7.1}, \ren{2-7.2}, and \ren{2-7.4} are cancellative.

Pseudomultiplications with $\htE=\sup E$, especially for $E=I$ (as in
Examples \ren{2-7.2} and \ren{2-7.3}), are called triangular norms, and
the semiring is then called absorptive.
\tme{Exa}{2-7}

\tma{Def}{2-8}
A function $f:X\rightarrow E$ is called measurable if for every $a\in E$
the set $ \{x\in X: f(x)\leq a \} $ is in ${\cal A}$.

The space of measurable functions $f:X\rightarrow E$ is denoted by
${\cal M}$.
\tme{Def}{2-8}

The definition of pseudointegrals of measurable functions is strictly
analogous to the definition of conventional integrals:

\tma{Def}{2-9}
For $A\subset X$ define the characteristic function
\[ \chi_A(x) = \left\{ {\htE {\quad \rm if} x\in A   \atop
                        \htN {\quad \rm if} x\not\in A}  \right. .   \]
A simple function is a function $h:X\rightarrow E$ which can be written
as
\[ h(x) = \htSu_{i=1}^{n} h_i \htm \chi_{A_i}  \]
with disjoint $A_i\in {\cal A}$, $h_i\in E$, $i=1,\dots ,n$ ($n\in
\Nset$).
We denote the space of measurable simple functions by ${\cal H}$.

For $h\in {\cal H}$, $B\in {\cal A}$, and a pseudomeasure $\htP$ we
define
\[ I(h;B;\htP) = \htSu_{i=1}^{n} h_i \htm \htP (A_i \cap B) .\]
\tme{Def}{2-9}

\tma{Def}{2-10}
The pseudointegral of a measurable function $f:X\rightarrow E$ over a
set $B\in {\cal A}$ with respect to a pseudomeasure $\htP$ is defined by
\[ \htIn_B\, f(x)\htm \htP(dx) = \sup \{I(h;B;\htP):h\in {\cal H},
   h \leq f\} .   \]
Here, the notation $h\leq f$ means that $h(x)\leq f(x)$ for all $x\in
X$.
\tme{Def}{2-10}

\tma{Exa}{2-10a}
\rm
In many situations there is a particularly simple reference
pseudomeasure $\htQ$ with respect to which most pseudointegrals are
calculated, in which case we use the following abbreviated notation:
\[ \htIn_X\, f(x)\htm \htQ(dx) = \htIn_X\, f(x)\htm dx  . \]

In the case $E=\Eset^+$, $\hta=+$, $\htN = 0$, and $X=\Rset^n$ the
standard reference measure $\htQ$ is the Lebesgue measure.

In the case $a\hta b = \sup\{a,b\}$ we use
\[ \htQ(A) = \htE  \mbox{ for all } A\in {\cal A} - \{ \emptyset \} . \]
Then we obtain
\[ \htIn_A\, f(x)\htm dx = \sup \{f(x):x\in A\}   . \]
\tme{Exa}{2-10a}

The first part of the following proposition is obvious, and the second
part can be proved like the theorem of monotone convergence for
conventional integrals:

\tma{Pro}{2-11}
\begin{enumerate}
 \item If $f\leq g$ for measurable $f$, $g$, then
       $\htIn_X\, f(x)\htm \htP(dx) \leq \htIn_X\, g(x)\htm \htP(dx)$.
 \item Let $\{f_i\}$ be a sequence of measurable functions with
       \[ f_1 \leq f_2 \leq \dots .\] Then
       \[ \lim_i \htIn_X\, f_i(x)\htm \htP(dx)
        = \htIn_X\, \lim_i f_i(x)\htm \htP(dx)  .\]
\end{enumerate}
\tme{Pro}{2-11}

Pseudointegrals are pseudolinear. Pseudolinearity is a generalisation of
linearity which involves the concept of semimodules as a generalisation
of vectorspaces.

\tma{Def}{2-12}
A positively ordered semimodule over a positively ordered semi\-ring
$(E,\leq,\hta,\htN,\htm,\htE)$ is a positively ordered commutative
monoid $(V,\leq,$ $\hta,\htN)$ for which an external operation,
$\htm: E\times V\rightarrow V$, called pseudomultiplication by a scalar,
is defined and has the following properties:
\begin{enumerate}
 \item $(a\htm b)\htm v = a \htm (b\htm v)$
 \item $(a\hta b)\htm v = (a \htm v) \hta (b \htm v)$
 \item $a \htm (v\hta w) = (a\htm v) \hta (a\htm w)$
 \item $\htN \htm v = \htN$
 \item $\htE \htm v = v$
 \item $a\leq b \Rightarrow a\htm v \leq b\htm v$
 \item $v\leq w \Rightarrow a\htm v \leq a\htm w$
\end{enumerate}
where $a,b\in E$ and $v,w\in V$.

Note that in this definition each of the symbols $\hta,\htm$, and $\htN$
has two different meanings which, however, can clearly be distinguished
by the context in which the symbols are used.
\tme{Def}{2-12}

Special cases of semimodules are modules, where pseudoaddition is
invertible, and vectorspaces, where additionally pseudomultiplication
is invertible for non-$\htN$ elements.

\tma{Def}{2-13}
Let $V$ and $W$ be semimodules over a semiring $E$. A map
$H:V\rightarrow W$ is called a semimodule morphism or a pseudolinear map
if
\begin{enumerate}
 \item $H(v\hta w) = H(v) \hta H(w)$
 \item $H(a\htm v) = a\htm H(v)$
\end{enumerate}
for all $v,w\in V$ and $a\in E$.
\tme{Def}{2-13}

Pseudolinearity of pseudointegration is expressed in the following
obvious pro\-position.

\tma{Pro}{2-14}
$({\cal M},\leq,\hta,\htN)$ with
\[  (f\hta g)(x) = f(x)\hta g(x) ,\]
\[  \htN(x) = \htN , \]
\[  (a\htm f)(x) = a\htm f(x)  \]
for all $f,g\in{\cal M}$, $x\in X$, and $a\in E$, and $\leq$ defined as
in Definition \rse{2-10} is a positively ordered semimodule over
$(E,\leq,\hta,\htN,\htm,\htE)$.

Pseudointegration defines a semimodule morphism
\begin{eqnarray*}
 I: {\cal M} & \rightarrow & E \\
          f  & \mapsto & \htIn_X\, f(x)\htm \htP(dx)
\end{eqnarray*}
\tme{Pro}{2-14}

Measurable functions can be used to define new pseudomeasures via
pseudointegrals.

\tma{Pro}{2-15}
If $f\in {\cal M}$, the set function $\htP_f:{\cal A}\rightarrow E$
defined by
\[ \htP_f(A) = \htIn_A\,f(x)\htm\htP(dx)  \]
is a pseudomeasure.
\tme{Pro}{2-15}

The function $f$ is then called the pseudodensity of $\htP_f$ with
respect to $\htP$, and $\htP_f$ is called absolutely continuous with
respect to $\htP$.

If $\htP_f$ is not absolutely continuous with respect to $\htP$ there
is no pseudodensity as a measurable function, but it may be
defined as a distribution.

\rsec{Uncertain dynamical systems \protect\\defined by transition
      pseudomeasures}{3}

The tool of pseudomeasures makes it possible to define a quite general
concept of uncertain dynamical systems with discrete time:

\tma{Def}{3-1}
An uncertain dynamical system with discrete time on a measurable phase
space $(X,{\cal A})$ is defined by a family $(\htP_x)_{x\in X}$ of
pseudomeasures on $X$ with values in a positively ordered commutative
semiring $(E,\leq,\hta,\htN,\htm,\htE)$.

The system is called normal if $\htP_x(X) = \htE$ for all $x\in X$.
\tme{Def}{3-1}

\tma{Rem}{3-2}
\rm
The pseudomeasures $\htP_x$ are interpreted as one-step transition
pseudomeasures: If at time $t$ the system is in a state $x$ then the
guess that at time $t+1$ the system will be in a state in $A\in {\cal
A}$ is assessed at $\htP_x(A)$.

This definition includes deterministic discrete-time systems, usually
defined by a map $F:X\rightarrow X$, as a special case, namely the case
$\htP_x = \htdel_{F(x)}$. Here, $\htdel_x$ is the Dirac pseudomeasure,
defined by
\[
 \htdel_x(A) = \left\{ {\htE \ \mbox{ if } x\in A     \atop
                        \htN \ \mbox{ if } x\not\in A     } \right.
\]
\tme{Rem}{3-2}

\tma{Def}{3-3}
For $N\in\Nset$ the $N$-step transition pseudomeasure from $x\in X$ to
$A\in{\cal A}$ is defined by the following recursion formula:
\[  \htP_x^{[N]}(A) = \htIn_X \, \htP_y(A) \htm \htP_x^{[N-1]}(dy) .\]
\tme{Def}{3-3}

\tma{Rem}{3-4}
\rm
Pseudoaddition decides about how one-step transitions to uni\-ons of sets
are assessed and is therefore related to a generalised logical ``or''.

Pseudomultiplication comes into play when evaluating multi-step
transitions. It is related to a generalised ``and''. For example, the
two-step transition measure is obtained by pseudointegrating over all
intermediate points the pseudoproduct of the assessment of doing first
one step to the intermediate point and then one step from there to the
final set.
\tme{Rem}{3-4}

\tma{Exa}{3-5}
\rm
The best-known example of an uncertain dynamical systems is a stochastic
dynamical system. In this case, $E=\Eset^+$, $\hta=+$, $\htm=\cdot$,
$\htN=0$, $\htE=1$ --- like in Example \rse{2-7}\ren{2-7.1} --- and
the transition pseudomeasures are transition probabilities. Stochastic
systems have been used extremely successfully for modelling
uncertainties. This is especially true in all situations where, at least
in principle, the frequencies of certain transitions can be observed at
a large number of identical copies of the system.

But there are situations where other rules of assessing uncertainties
can be more useful. Here is an example: Assume that changing a system
from state $x$ into state $y$ costs an amount $-p(x,y)$ where $p(x,y)\in
\Eset^-$ (we count the costs as negative gains). We do not know who runs
the system, so we cannot predict its future with certainty, but we
assume that whoever runs the system will try to minimise the costs. So a
reasonable assessment of the guess that the system is in a state in $A$
at time $t+1$ after being in state $x$ at time $t$ is
\[ \htP_x(A) = \sup \{p(x,y):y\in A \} , \]
and a guess for being in $A$ at time $t+2$ would be assessed at
\[ \htP_x^{[2]}(A) = \sup \{p(x,y)+p(y,z):y\in X, z\in A \} . \]
This means that for this example we would choose $E=\Eset^-$,
$\hta=\sup$, $\htm=+$, $\htN=-\infty$, and $\htE=0$ --- like in Example
\rse{2-7}\ren{2-7.4}.

This example shows that the choice $\hta=\sup$ is typical for systems
which can be formulated as optimisation problems.

In the context of fuzzy set theory the choice $\hta=\sup$ is often
denoted by the adjective ``possibilistic''. For instance, the
pseudomeasures for that choice are called possibility measures.
They offer a more diverse assessment of the possibility of events
than a strict Boolean classification of states into possible and
impossible states. The Boolean case is realised by the semiring
$(\{0,1\},$ $\leq,\sup,0,\inf,1)$. Replacing the evaluation set $\{0,1\}$
by the unit interval $I$, one can try to quantify colloquial expressions
like ``nearly impossible'' or ``maybe possible''.
\tme{Exa}{3-5}

\tma{Def}{3-6}
A pseudomeasure $\htS^*$ is called invariant pseudomeasure of the
uncertain system $(\htP_x)$ if for all $A\in{\cal A}$
\[  \htS^*(A) = \htIn_X \, \htP_y(A) \htm \htS^*(dy) .\]
\tme{Def}{3-6}

\tma{Rem}{3-7}
\rm
Invariant pseudomeasures of uncertain systems are important for their
long-term behaviour. If we assess the presumable state of a system at
time $t$ on the basis of a pseudomeasure $\htS_t$ then the pseudomeasure
$\htS_{t+1}$, defined by
\[  \htS_{t+1}(A) = \htIn_X \, \htP_y(A) \htm \htS_t(dy) \]
for $A\in{\cal A}$, gives the assessment at time $t+1$.

Therefore invariant pseudomeasures are fixed points of the dynamics of
assessments, and depending on their stability properties they may
characterise the importance of subsets of the state space for the
presumable long-term behaviour of the uncertain system.

Normality of a system guarantees that $\htS_{t+1}(X) = \htS_t(X)$,
indicating that the system is closed.
\tme{Rem}{3-7}

It is often convenient to work with pseudodensities instead of
pseudomeasures.

\tma{Def}{3-8}
Let $\htQ$ be a standard reference pseudomeasure (see Example
\rse{2-10a}). If the transition pseudomeasures $\htP_x$ have
pseudodensities $\htPp(x,\cdot)$, i. e.
\[
 \htP_x(A) = \htIn_A \htPp(x,y) \htm dy
\]
then these pseudodensities are called transition pseudodensities of the
uncertain dynamical system.
\tme{Def}{3-8}

In terms of transition pseudodensities, normality of the system means
\[
 \htIn_X \htPp(x,y) \htm dy = \htE
\]
for all $x\in X$.

\tma{Def}{3-9}
An uncertain dynamical system with transition pseudodensities 
$\htPp(x,y)$ is called deterministically motivated if there is a map
$F:X\rightarrow X$ such that
$ \htPp(x,y) \leq \htPp(x,F(x))  $
for all $x,y\in X$ and
$ \htPp(x,y) = \htPp(x,F(x)) $
only if $y=F(x)$.
\tme{Def}{3-9}

In such a system a transition from $x$ to $F(x)$ is assessed at the
highest value among all transitions so that this case models what
happens to the deterministic system given by $F$ if it is perturbed by
uncertainties.

\tma{Pro}{3-10}
If an uncertain system has transition pseudodensities \linebreak $\htPp(x,y)$
and 
if there is a function $\htSs\in{\cal M}$ which solves for all $y\in X$
the equation
\[
 \htSs(y) = \htIn_X \htSs(x)\htm\htPp(x,y)\htm dx
\]
then $\htP_{\htSs}$ (see Proposition \rse{2-15} for the notation) is an
invariant pseudomeasure. $\htSs$ is called invariant pseudodensity.

The dynamics of pseudomeasures described in Remark \rse{3-7} reads on
the level of pseudodensities as follows:
\[
 \htSs_{t+1}(y) = \htIn_X \htSs_t(x)\htm\htPp(x,y)\htm dx .
\]
\tme{Pro}{3-10}

\tma{Def}{3-11}
A measurable function $k:X\times X \rightarrow E$ defines a pseudolinear
operator $O_k$ on ${\cal M}$ by
\[
 (O_k\,f)(x) = \htIn_X k(x,y)\htm f(y)\htm dy .
\]
The function $k$ is called the pseudointegral kernel of $O_k$.

The transposed operator $O_k^*$ is defined by
\[
 (O_k^*\,f)(x) = \htIn_X f(y)\htm k(y,x) \htm dy .
\]
\tme{Def}{3-11}

\tma{Cor}{3-12}
In terms of the newly introduced operators, the equations in Proposition
\rse{3-10} read as follows:
\[ \htSs_{t+1} = O_{\htPp}^*\, \htSs_t  \]
and
\[ \htSs = O_{\htPp}^*\, \htSs .  \]
\tme{Cor}{3-12}

Generalising a further concept of linear algebra, we can say the
last equation means that the invariant pseudodensity $\htSs$ is an
eigenfunction of the operator $O_{\htPp}^*$ with eigenvalue $\htE$.

\tma{Def}{3-13}
Let $V$ be a semimodule over a commutative semiring $E$ and
$H:V\rightarrow V$ a pseudolinear map. If there is a $v\in V-\{\htN\}$
and an $a\in E$ such that
\[ H(v) = a\htm v \]
then $a$ is called an eigenvalue of $H$, and $v$ is called an
eigenelement corresponding to that eigenvalue. The set of eigenelements
corresponding to $a$ together with $\htN\in V$ is denoted by $V_a$.
\tme{Def}{3-13}

\tma{Pro}{3-14}
The space $V_a$ is a sub-semimodule of $V$.
\tme{Pro}{3-14}

Thus, the search for invariant pseudodensities of uncertain systems
is an eigen\-value-eigenfunction problem (or eigenproblem for short) on a
function space with the structure of a semimodule.

The special case of a stochastic system leads to a classical
eigenproblem on a vector space, and this is the situation which has
been studied most.

It is too much to expect that a lot can be said about the general
eigenproblem, but the vector space case is not the only one which can be
analysed in great detail. The following section deals with the case of
idempotent pseudoaddition, which is in some aspects even simpler than
the vector space case.

\rsec{The eigenproblem for pseudointegral \protect\\operators on
      idempotent semimodules}{4}

In order to construct eigenfunctions of pseudointegral operators on
semimodules it is useful to introduce iterated kernels and
transitive closures:

\tma{Def}{4-1}
Let $k(x,y)=k_1(x,y)$ be a pseudointegral kernel. The iterated kernels
are then defined for $N\geq 1$ by
\[ k_{N+1}(x,y) = \htIn_X k_N(x,z) \htm k(z,y) \htm dz . \]
In the case $k(x,y)=\htPp(x,y)$ they are called $N$-step transition
pseudodensities.

The transitive closure of the kernel is
\[ T_k(x,y) = \htSu_{N=1}^{\infty} k_N(x,y)   .  \]
\tme{Def}{4-1}

\tma{Def}{4-2}
A point $a\in X$ is called a basis point for a kernel $k(x,y)$ if there
is a (not necessarily unique) $c(a)\in E$, $c(a)\neq\htN$, such that
\[
 c(a)\htm T_k(a,a) =
 c(a)\htm (T_k(a,a) \hta \htdel(a,a)) 
\]
where $\htdel(y,x)$ is the pseudodensity of the Dirac pseudomeasure
$\htdel_x$ introduced in Remark \rse{3-2}.

The set of all basis points for $k(x,y)$ is denoted by $B_k$.
\tme{Def}{4-2}

\tma{Pro}{4-3}
Let $k(x,y)$ be a pseudointegral kernel. Then for every $a\in B_k$ and
$c(a)$ fulfilling the defining property of $B_k$, the function
$\Psi_{k,a}(x) = c(a)\htm T_k(x,a)$ is an eigenfunction of $O_k$ and the
function $\Phi_{k,a}(x) = c(a)\htm T_k(a,x)$ is an eigenfunction of
$O_k^*$, both with eigenvalue $\htE$.
\tme{Pro}{4-3}

\Proo
From
\[
  T_k(x,a) = \htIn_X k(x,y)\htm (T_k(y,a)\hta\htdel(y,a)) \htm dy
\]
we obtain by pseudomultiplying both sides by $c(a)$
\[
  \Psi_{k,a}(x) = \htIn_X k(x,y)\htm \Psi_{k,a}(y) \htm dy
                = O_k \Psi_{k,a}(x) .
\]
The statement about $\Phi_{k,a}$ can be proved analogously.
\Prooe

Proposition \rse{4-3} is obviously not useful in the conventional vector
space case, but it is the key to the eigenproblem in so called
idempotent semimodules.

\tma{Def}{4-4}
A semimodule $V$ is called idempotent if the pseudoaddition on the
underlying semiring $E$ is idempotent, i. e.,
\[ a\hta a = a \mbox{ for all } a\in E  .\]
\tme{Def}{4-4}

\tma{Rem}{4-5}
\rm
In a positively ordered monoid $E$ there is only one operation which
qualifies as an idempotent pseudoaddition: We know that for $a,b\in E$,
$a\leq a\hta b$ and $b\leq a\hta b$. Now assume $c\in E$ is another
upper bound of $a$ and $b$: $a\leq c$ and $b\leq c$. Then we have $a\hta
b \leq c\hta c = c$, showing that $a\hta b = \sup\{a,b\}$. Consequently,
the standard pseudointegral of a function is its supremum (cf. Example
\rse{2-10a}), and the Dirac pseudodensity $\htdel(a,a)$ appearing in
Definition \rse{4-2} is equal to $\htE$.

Thus, the case of idempotent semimodules is what we are interested
in when we study the possibilistic systems of Example \rse{3-5}.
\tme{Rem}{4-5}

\tma{Rem}{4-6}
\rm
In the case of idempotent semimodules the condition for $a\in X$ being a
basis point has an especially simple form if $k(x,y)$ is bounded above
by $\htE$. Since $T_k(x,y)$ is then bounded above by $\htE$, too, the
condition in Definition \rse{4-2} can be reduced to
\[
 c(a)\htm T_k(a,a) = c(a) .
\]
\tme{Rem}{4-6}

In a cancellative semiring this is obviously equivalent to
\[
 T_k(a,a) = \htE .
\]

\tma{Def}{4-7}
The set of normal basis points for a kernel $k(x,y)$ is defined as
\[ R_k = \{ x\in X: T_k(x,x) = \htE \}  .      \]
\tme{Def}{4-7}

$R_k$ is a subset of $B_k$. For $k$ bounded by $\htE$ in a
cancellative semiring we know $B_k=R_k$, but in the noncancellative case
this is generally not true.

\tma{Exa}{4-8}
\rm
In the case $a\htm b = \inf\{a,b\}$ it is easy to see that $B_k=X$.
For every $a\in X$ the condition of Definition \rse{4-2} can be
fulfilled with $c(a)=T_k(a,a)$.
\tme{Exa}{4-8}

Nevertheless, for kernels bounded by $\htE$ the normal basis points
have a special importance in any case:

\tma{Rem}{4-9}
\rm
The eigenfunctions from Proposition \rse{4-3} have the property
\linebreak
$\htIn_X \Psi_{k,a}(x) \htm dx = \htE$ and $\htIn_X \Phi_{k,a}(x)
\htm dx = \htE$ if and only if $c(a) = \htE$ so that $a\in R_k$. This
follows from the fact that $\Psi_{k,a}(x) \leq c(a) = \Psi_{k,a}(a)$
which means $\htIn_X \Psi_{k,a}(x) \htm dx = c(a)$.
\tme{Rem}{4-9}

Next, we study the case that $R_k$ contains more than one point and the
relation between the different eigenfunctions that can then be
constructed as described in Proposition \rse{4-3}.

\tma{Def}{4-10}
Two normal basis points $a,b\in R_k$ are called equivalent $a\sim_k b$
if
\[ T_k(a,b) \htm T_k(b,a) = \htE . \]
Since the relation $\sim_k$ is an equivalence relation on $R_k$, every
normal basis point $a\in R_k$ is a representative of an equivalence
class, which is denoted by $[a]_k$.
\tme{Def}{4-10}

\tma{Pro}{4-11}
If $a\sim_k b$ for $a,b\in R_k$ then
\[ \Phi_{k,a}(x) = T_k(a,b) \htm \Phi_{k,b}(x)  \]
and
\[ \Psi_{k,a}(x) = T_k(b,a) \htm \Psi_{k,b}(x) . \]
\tme{Pro}{4-11}

\Proo
We show the first statement; the second follows analogously.

We have
\begin{eqnarray*}
  T_k(a,b) \htm \Phi_{k,b}(x) & \leq & \htIn_X T_k(a,y)
                                     \htm T_k(y,x) \htm dy
  =
  \htSu_{N,M=1}^{\infty} k_{N+M}(a,x) \\
  =                         &  & \htSu_{N=2}^{\infty} k_N(a,x)
                             \leq \Phi_{k,a}(x) .
\end{eqnarray*}

On the other hand, from equivalence of $a$ and $b$ follows
\begin{eqnarray*}
  \Phi_{k,a}(x) & = & T_k(a,b) \htm T_k(b,a) \htm \Phi_{k,a}(x) \\
                & \leq & T_k(a,b) \htm \Phi_{k,b}(x)  ,
\end{eqnarray*}
which completes the proof.
\Prooe

Proposition \rse{4-11} shows that the eigenfunctions constructed with
equivalent normal basis points differ by a scalar factor only.

In order to derive more results about the eigenproblem we need more
assumptions about the space $X$ and the kernel $k$.

We assume in the following that the state space $X$ is a compact metric
space.

It will turn out to be useful to have sort of a mean-value theorem for
pseudointegrals, and this will dictate the regularity property we
require of functions that appear as pseudointegrands.

\tma{Def}{4-12}
A function $f:X\rightarrow E$ is called upper semicontinuous if for all
$x\in X$ and for all sequences $(x_i)$ with
$\lim_{i\rightarrow\infty}x_i=x$
\[ \overline{\lim} f(x_i) \leq f(x)  .  \]
\tme{Def}{4-12}

\tma{Lem}{4-13}
Let $X$ be a compact metric space and $f:X\rightarrow E$
upper semicontinuous. Then under each one of the conditions
 \begin{itemize}
   \item[(i)]  $(E,\leq)$ is totally ordered,
   \item[(ii)] $X$ is pathwise connected,
 \end{itemize}
there exists a $y\in X$ such that
\[
   f(y) = \htIn_X f(x) \htm dx .
\]
\tme{Lem}{4-13}

\Proo
Under the first condition the result follows like the well known result
that upper semicontinuous real functions attain their
suprema on compact sets.

The second condition is sufficient since all elements in the image of a
path in $X$ under an upper semicontinuous function are comparable.
\Prooe

From now on we make assumptions that guarantee that Lemma \rse{4-13} can
be applied. In particular we assume that $k$ is upper semicontinuous in
both arguments (which implies that $k_N$ and $T_k$ are upper
semicontinuous) and that either all elements of $E$ are comparable or
all sets over which integrals are taken are pathwise connected.

Proposition \rse{4-3} can be used to show that $\htE$ is an eigenvalue
and to construct corresponding eigenfunctions only if $B_k$ is not
empty. The following definition leads to a sufficient condition for
existence of a normal basis point.

\tma{Def}{4-14}
The kernel $k$ is called definite if $\htIn_X T_k(x,x) \htm dx = \htE$.
\tme{Def}{4-14}

A direct consequence of Lemma \rse{4-13} is

\tma{Pro}{4-15}
Under the conditions of Lemma \rse{4-13} and if $k$ is definite then
$R_k\neq\emptyset$.
\tme{Pro}{4-15}

An important class of definite kernels are the normal kernels.

\tma{Pro}{4-16}
A normal kernel $k$ is bounded by $\htE$ and definite.
\tme{Pro}{4-16}

\Proo
Suppose that there are $x,y\in X$ such that $k(x,y)$ is not less or
equal $\htE$. Then clearly $\htIn_X k(x,z) \htm dz > \htE$ in
contradiction of the normality of $k$. So $k(x,y)$ is bounded by $\htE$
for all $x,y \in X$.

Since $k$ is bounded by $\htE$, so are the iterated kernels $k_N$
($N\in\Nset$) and $T_k$.

Normality of $k$ and Lemma \rse{4-13} imply that for every $x\in X$
there
is a $y\in X$ such that $k(x,y)=\htE$. Therefore we can find an infinite
sequence $(x_i)_{i=1,2,\dots}$ such that for every $n\in \Nset$
\[ k_n(x_1,x_{n+1}) = \htE .  \]
Since $X$ is a compact metric space there is a subsequence
$(z_j)=(x_{i_j})$ which converges to some $z\in X$. For any $m,n\in
\Nset$ with $m<n$ we have
\[ T_k(z_m,z_n) \geq k_{i_n-i_m}(z_m,z_n) = \htE , \]
and therefore $T_k(z_m,z_n)=\htE$.

Now fix $m$ and let $n\rightarrow\infty$. Upper semicontinuity of $T_k$
in the second argument implies $T_k(z_m,z) \geq \htE$ and therefore
$T_k(z_m,z)=\htE$. Finally, $m\rightarrow\infty$ and upper
semicontinuity in the first argument gives $T_k(z,z)=\htE$ which shows,
together with the upper bound $\htE$ for $T_k$, that $k$ is definite.
\Prooe

The next statement shows that eigenfunctions constructed like in
Proposition \rse{4-3} are pseudolinearly independent if they start from
nonequivalent normal basis points.

\tma{Pro}{4-17}
If $K$ is a compact subset of $R_k$, $k$ a kernel bounded by $\htE$,
$\lambda$ an upper semicontinuous function on $K$, and for $b\in R_k$
and all $x\in X$
\[
\Phi_{k,b}(x) = \htIn_K \lambda(a) \htm \Phi_{k,a}(x) \htm da ,
\]
then $b$ is equivalent to one of the elements of $K$.

An analogous result holds for the functions $\Psi_{k,a}(x)$.
\tme{Pro}{4-17}

\Proo
We have
\[
\htE = \Phi_{k,b}(b) = \htIn_K \lambda(a) \htm \Phi_{k,a}(b) \htm da .
\]
By Lemma \rse{4-13} there is an $\alpha\in K$ such that
\[ \htE = \lambda(\alpha) \htm \Phi_{k,\alpha}(b) . \]
Since $k$ is bounded by $\htE$ we have $\Phi_{k,a}(x)\leq\htE$ for all
$a\in R_k$ and all $x\in X$. Therefore the last equation shows that
$\lambda(\alpha)$ cannot be smaller than $\htE$. But on the other hand
we have
\[ \htE \geq \Phi_{k,b}(\alpha) \geq \lambda(\alpha) \htm
             \Phi_{k,\alpha}(\alpha) = \lambda(\alpha)    \]
from which we can conclude $\lambda(\alpha) = \htE$. But this means that
\[ \Phi_{k,\alpha}(b) = \Phi_{k,b}(\alpha) = \htE \]
which shows that $b\sim_k \alpha$.
\Prooe

The next result shows that the construction of Proposition \rse{4-3}
leads to all eigenfunctions of $O_k$ and $O_k^*$ with eigenvalue $\htE$.

\tma{Pro}{4-18}
Let $k$ be a kernel bounded above by $\htE$ and $\varphi$ a
positive, upper semicontinuous eigenfunction of $O_k^*$ with eigenvalue
$\htE$. Then
\[ \varphi(x) = \htIn_{B_k} \varphi(a) \htm T_k(a,x) \htm da  \]
for all $x\in X$.

Similarly, for a positive, upper semicontinuous eigenfunction $\psi$ of
$O_k$ with eigenvalue $\htE$:
\[ \psi(x) = \htIn_{B_k} \psi(a) \htm T_k(x,a) \htm da  \]
for all $x\in X$.
\tme{Pro}{4-18}

\Proo
If $\varphi$ is an eigenfunction of $O_k^*$ with eigenvalue $\htE$ it is
an eigenfunction of $O_{T_k}^*$ with eigenvalue $\htE$, too, and therefore:
\[ \varphi(x) = \htIn_X \varphi(a) \htm T_k(a,x) \htm da
           \geq \htIn_{B_k} \varphi(a) \htm T_k(a,x) \htm da   \]
for all $x\in X$.

On the other hand we can use the eigenfunction property of $\varphi$ and
Lemma \rse{4-13} to construct a sequence $(x_i)$ such that for all $i$
\[ \varphi(x_i) = \varphi(x_{i+1}) \htm k(x_{i+1},x_i)   \]
and therefore for $j>i$
\begin{eqnarray*}
   \varphi(x_i) & = & \varphi(x_j) \htm k(x_j,x_{j-1})
                       \htm \dots \htm k(x_{i+1},x_i)    \\
                & \leq & \varphi(x_j) \htm k_{(j-i)}(x_j,x_i)   \\
                & \leq & \varphi(x_j) \htm T_k(x_j,x_i)  .
\end{eqnarray*}

We start the construction of $(x_i)$ with $x_0=x\in X$. The sequence
$(x_i)$ has a convergent subsequence $(x_{i_n})$ with
$\lim_{n\rightarrow\infty} x_{i_n} = b \in X$, and we will show later
that $b\in B_k$. Using the upper semicontinuity of $\varphi$ and $T_k$
we obtain in the limit $n\rightarrow\infty$
\begin{eqnarray*}
  \varphi(x) & \leq & \overline{\lim} \varphi(x_{i_n}) \htm
                      T_k(x_{i_n},x) \\
             & \leq & \varphi(b) \htm T_k(b,x) \\
             & \leq & \htIn_{B_k} \varphi(a) \htm T_k(a,x) \htm da
\end{eqnarray*}
which together with the opposite inequality establishes the statement about
$\varphi$.

The last step is to show that $b\in B_k$. We use one of the above
inequalities:
\[
  \varphi(x_{i_m}) \leq \varphi(x_{i_n}) \htm T_k(x_{i_n},x_{i_m})
\]
and let first $n\rightarrow\infty$ and then $m\rightarrow\infty$. Using
the upper semicontinuity of $T_k$ we obtain
\[
  \overline{\lim} \varphi(x_{i_n}) \leq
  \overline{\lim} \varphi(x_{i_n}) \htm T_k(b,b) .
\]
But since $T_k$ is bounded above by $\htE$ this means
\[
  \overline{\lim} \varphi(x_{i_n}) =
  \overline{\lim} \varphi(x_{i_n}) \htm T_k(b,b) .
\]
Therefore, for $b$ the condition of Definition \rse{4-2} is fulfilled
with $c(b)=\overline{\lim} \varphi(x_{i_n})$.
\Prooe

Now we turn to the question of eigenvalues different from $\htE$.

\tma{Def}{4-19}
A kernel $k$ is called strongly connected if for all $x,y\in X$, $x\neq
y$,
\[ T_k(x,y) > \htN . \]
\tme{Def}{4-19}

\tma{Rem}{4-20}
\rm
A kernel is certainly strongly connected if $k(x,y)>\htN$ for all
$x,y\in X$, $x\neq y$. A kernel is certainly not strongly connected if
there are two nonempty sets $X_1, X_2=X-X_1$ such that $k(x,y)=\htN$ for
all $x\in X_1$ and $y\in X_2$, because for those $x,y$ it is clear that
$T_k(x,y)=\htN$.
\tme{Rem}{4-20}

\tma{Pro}{4-21}
If $k$ is strongly connected and if $\psi$ is an upper semicontinuous
eigenfunction of $O_k$ (or $O_k^*$) with eigenvalue $\lambda$ then
$\lambda > \htN$ and $\psi(x) > \htN$ for all $x\in X$.
\tme{Pro}{4-21}

\Proo
Assume that $\lambda = \htN$. Then the eigenvalue equation implies
\[
  k(x,y) \htm \psi(y) = \htN
\]
for all $x,y\in X$. There must be at least one $y^*\in X$ such that
$\psi(y^*)\neq \htN$. This means $k(x,y^*)=\htN$ for all $x\in X$
in contradiction to the strong connectedness of $k$. Therefore $\lambda
> \htN$.

Assume now that $X_1=\{x\in X:\psi(x)=\htN\}$ is not empty. $X_2=X-X_1$
also is not empty. The eigenvalue equation leads to
\[
  \htIn_{X_2} k(x,y) \htm \psi(y) \htm dy = \htN
\]
for all $x\in X_1$, and this implies
\[
  k(x,y) = \htN
\]
for all $x\in X_1$ and $y\in X_2$. Remark \rse{4-20} shows that this is
a contradiction to the strong connectedness of $k$; hence $X_1$ must be empty.
\Prooe

\tma{Pro}{4-22}
If $k$ is a definite, strongly connected kernel with values in a
cancellative semiring and $\lambda$ an eigenvalue of $O_k$ or $O_k^*$
with an upper semicontinuous eigenfunction $\psi$ then
$\lambda=\htE$.
\tme{Pro}{4-22}

\Proo
First we show that $\lambda < \htE$ is not possible.

The eigenvalue equation implies that
\[
 \lambda \htm \psi(x) \geq k(x,y) \htm \psi(y)
\]
for all $x,y\in X$. Consider a cyclic sequence $y_0, y_1, \dots,
y_N=y_0$. Repeated use of the above inequality shows
\[
 k(y_0,y_1)\htm\dots\htm k(y_{N-1},y_0) \leq \lambda\htm\dots\htm\lambda
 \leq \lambda
\]
where we used already the assumption $\lambda<\htE$ and the cancellation
law, keeping in mind that $\psi$ is positive.

This is true for any $y_0\in X$ and any cyclic sequence from $y_0$ to
$y_0$. Taking the supremum we find
\[
 \htIn_X T_k(y,y) \htm dy \leq \lambda < \htE
\]
which contradicts the definiteness of $k$.

Next we show that the assumption $\lambda > \htE$ leads to a
contradiction, too.

Construct a sequence $(x_i)$ such that
\[
 \lambda \htm \psi(x_i) = k(x_i,x_{i+1}) \htm \psi(x_{i+1})  .
\]
For $n>m$ this leads to
\[
 \lambda \htm \psi(x_m) \leq T_k(x_m,x_n) \htm \psi(x_n) .
\]
Concentrating on a convergent subsequence and using semicontinuity this
implies
\[
 \htE < \lambda \leq T_k(b,b)
\]
for $b$ an accumulation point of $(x_i)$. But this contradicts the
definiteness of $k$ again.
\Prooe

\tma{Pro}{4-23}
If $k$ is a strongly connected kernel with values in a semi\-ring with
multiplicative inverses and $\lambda$ an eigenvalue of $O_k$ or $O_k^*$
with an upper semicontinuous eigenfunction $\psi$ then the kernel
$\lambda^{-1}\htm k(x,y)$ is definite.
\tme{Pro}{4-23}

\Proo
The eigenvalue equation for $O_k$ allows the construction of a sequence
$(x_i)$ such that
\[
  \psi(x_i) = \lambda^{-1} \htm k(x_i,x_{i+1}) \htm \psi(x_{i+1}) .
\]
Introducing the abbreviation $\tilde{k}(x,y)=\lambda^{-1}\htm k(x,y)$
we obtain
\[
  \psi(x_m) \leq T_{\tilde{k}}(x_m,x_n) \htm \psi(x_n)
\]
for $m<n$. For an accumulation point $b$ of $(x_i)$ this means
\[
  \htE \leq T_{\tilde{k}}(b,b) .
\]
On the other hand the eigenvalue equation can be used to show the
estimate
\[
  \htIn_X T_{\tilde{k}}(x,x) \htm dx \leq \htE .
\]
Both results together lead to
\[
  \htIn_X T_{\tilde{k}}(x,x) \htm dx = \htE .
\]
\Prooe

\tma{Rem}{4-24}
\rm
Proposition \rse{4-23} shows that in the case of a strongly connected
kernel and under the assumption of multiplicative invertibility the
eigenvalue is unique, and the eigenfunctions can be found by studying a
definite kernel.

The fact that
\[
  \htIn_X T_{\tilde{k}}(x,x) \htm dx = \htE
\]
leads to an interesting interpretation of the eigenvalue $\lambda$, at
least in a radicable semiring. A semiring $E$ is called radicable if
for every $b\in E$ and every $n\in \Nset$ the equation
\[ \htPr_{i=1}^{n} a = b \]
has a unique solution $a\in E$, denoted by $\htRt{n}{b}$.

The definiteness of $\tilde{k}$ means that the supremum over all
products \linebreak $\htPr_{i=1}^{N} \lambda^{-1} \htm k(x_{i-1},x_i)$ over
cyclic sequences with $x_0=x_N$ is $\htE$, or, if we call
\[ \htRt{N}{\htPr_{i=1}^{N} k(x_{i-1},x_i)} \]
a cycle mean, that $\lambda$ is the maximal cycle mean.
\tme{Rem}{4-24}

We end this section with an example which shows that the cancellation
law was essential for the derivation of uniqueness of the eigenvalue.
Here is the extreme non-cancellative situation of Example \rse{4-8}
again:

\tma{Exa}{4-25}
\rm
In the case $a\htm b = \inf\{a,b\}$ every $\lambda\in E$,
$\lambda\neq\htN$, is an eigenvalue of every $O_k$ and $O_k^*$. We know
from Example \rse{4-8} that $\htE$ is an eigenvalue. So there exists a
function $\psi$ such that
\[ \htIn_X k(x,y) \htm \psi(y) \htm dy = \psi(x)  .  \]
But then $\lambda\htm\psi$ fulfills for any $\lambda\neq\htN$ the
eigenvalue equation
\[ \htIn_X k(x,y) \htm \lambda \htm \psi(y) \htm dy
   = \lambda \htm \lambda \htm \psi(x)  ,  \]
showing that $\lambda$ is an eigenvalue.
\tme{Exa}{4-25}

\rsec{Connections between possibilistic \protect\\ and
      deterministic systems}{5}

In Remark \rse{3-7} and Corollary \rse{3-12} we noted already that
the eigenfunctions $\varphi$ of a pseudointegral operator $O^*_{\htPp}$
on an idempotent semimodule with eigenvalue $\htE$, which we found in
the previous section, are important for the long term behaviour of a
possibilistic system with transition possibility density $\htPp$: They
are the fixed points of a dynamics of possibility densities. If at some
moment the possibility of finding the system in the state $x$ is
assessed at $\varphi(x)$ for every $x\in X$, then this assessment will
not change in the future.

Now it is interesting what will happen to initial possibility
densities which are not eigenfunctions of $O^*_{\htPp}$ --- whether
they converge to one of the stationary possibility densities.
Unfortunately, such a convergence is not guaranteed in general, but only
in special cases.

\tma{Pro}{5-1}
Let a possibilistic system on a compact metric space $X$ be described by
a transition possibility density $\htPp(x,y)$ which is upper
semicontinuous and bounded above by $\htE$. If $\htPp(a,a)=\htE$ for
every $a\in B_{\htPp}$ then in the dynamics of possibility densities,
\[
  \htSs_{t+1} = O^*_{\htPp} \htSs_t ,
\]
starting from any initial possibility density $\htSs_0\in{\cal M}$,
$\htSs_t$ converges pointwise to an invariant possibility density as
$t\rightarrow\infty$.

\tme{Pro}{5-1}

\Proo
All we have to do is to show that
$\lim_{N\rightarrow\infty} \htPp_N(x,y)$ exists for all $x,y \in X$
since then
\[
  \lim_{t\rightarrow\infty} \htSs_t(y) = \htIn_X \htSs_0(x) \htm
  \lim_{t\rightarrow\infty} \htPp_t(x,y) \htm dx
\]
and
\[
  \htIn_X \lim_{t\rightarrow\infty} \htSs_t(x) \htm \htPp(x,y) \htm dx
  = \htIn_X \htSs_0(x) \htm \lim_{t\rightarrow\infty} \htPp_{t+1}(x,y)
  \htm dx = \lim_{t\rightarrow\infty} \htSs_t(y) .
\]

We will show that for all $x,y\in X$ there is a basis point $a\in
B_{\htPp}$ such that
\[
  \lim_{N\rightarrow\infty} \htPp_N(x,y) =
            T_{\htPp}(x,a) \htm T_{\htPp}(a,y) .
\]

Consider an infinite sequence $(\xi_i)$ in $X$ such that
\[
  \overline{\lim_N}\ \htPp_N(x,y) =
  \lim_{n\rightarrow\infty} \htPp(x,\xi_1) \htm \htPr_{j=1}^{n-1}
                            \htPp(\xi_j,\xi_{j+1}) \htm \htPp(\xi_n,y)
  .
\]
There is a subsequence $(\xi_{i_k})$ converging to some $a\in X$. We
will show now that $a$ is a basis point.

For any $k$ we have
\[
  \htPr_{j=i_k}^{i_{k+1}} \htPp(\xi_j,\xi_{j+1})   =
  \htPp_{i_{k+1}-i_k}(\xi_{i_k},\xi_{i_{k+1}})
\]
and therefore
\[
  \overline{\lim_N}\ \htPp_N(x,y)
\]
\[
  = \htPp_{i_{k+1}-i_k}(\xi_{i_k},\xi_{i_{k+1}}) \htm
  \lim_{n\rightarrow\infty}
  \htPp(x,\xi_1) \htm \dots \htm \htPp(\xi_{i_k -1},\xi_{i_k})
  \htm \htPp(\xi_{i_{k+1}},\xi_{i_{k+1}+1}) \htm \dots \htm
  \htPp(\xi_n,y)
\]
\[
  \leq
  T_{\htPp}(\xi_{i_k},\xi_{i_{k+1}}) \htm
  \lim_{n\rightarrow\infty}
  \htPp(x,\xi_1) \htm \dots \htm \htPp(\xi_{i_k -1},\xi_{i_k})
  \htm \htPp(\xi_{i_{k+1}},\xi_{i_{k+1}+1}) \htm \dots \htm
  \htPp(\xi_n,y) .
\]
Taking the upper limit over $k$ and using upper semicontinuity we see
that
\[
  \overline{\lim_N}\ \htPp_N(x,y)
\]
\[
  \leq
  T_{\htPp}(a,a) \htm
  \left[
  \overline{\lim_{k}}\lim_{n\rightarrow\infty}
  \htPp(x,\xi_1) \htm \dots \htm \htPp(\xi_{i_k -1},\xi_{i_k})
  \htm \htPp(\xi_{i_{k+1}},\xi_{i_{k+1}+1}) \htm \dots \htm
  \htPp(\xi_n,y) \right] .
\]
But the factor in square brackets cannot be larger than
$\overline{\lim_N}\ \htPp_N(x,y)$. Together with
$T_{\htPp}(a,a)\leq\htE$ this implies
\[
  \overline{\lim_N}\ \htPp_N(x,y)
  = T_{\htPp}(a,a) \htm \overline{\lim_N}\ \htPp_N(x,y) .
\]
According to Remark \rse{4-6} with
$c(a)=\overline{\lim_N}\ \htPp_N(x,y)$ this means that $a\in
B_{\htPp}$.

Now we know on the one hand that,
\[
  \overline{\lim_N}\ \htPp_N(x,y) \leq T_{\htPp}(x,a) \htm
                                              T_{\htPp}(a,y)
\]
and on the other hand that
\[
  \underline{\lim}_N\ \htPp_N(x,y) \geq
  \underline{\lim}_N\ \htPp_N(x,a) \htm
  \underline{\lim}_N\ \htPp_N(a,y) .
\]

The assumption $\htPp(a,a)=\htE$ guarantees that
\[
\ds \lim_{N\rightarrow\infty} \htPp_N(x,a) = T_{\htPp}(x,a)
\mbox{ and }
\ds \lim_{N\rightarrow\infty} \htPp_N(a,y) = T_{\htPp}(a,y)  ,
\]
and therefore
\[
  \lim_{N\rightarrow\infty} \htPp_N(x,y) =
            T_{\htPp}(x,a) \htm T_{\htPp}(a,y) .
\]
\vspace{-1ex}

\Prooe

The previous proposition shows again that the basis points of a definite
possibilistic system are decisive for its long-term behaviour. It is
interesting to look for the meaning of basis points in a normal
deterministically motivated system, i. e. according to Definition
\rse{3-1} and Definition \rse{3-9} a system with $\htPp(x,y)=\htE$ if
and only if $y=F(x)$ where the map $F:X\rightarrow X$ describes a
deterministic system.

We first recall some concepts from the theory of deterministic systems
(see e.g.\cite{Ru1}).

\tma{Def}{5-2}
Let a deterministic discrete time dynamical system be defined by a
continuous map $F:X\rightarrow X$ on a metric space $X$.

The set of non-wandering points for $F$, denoted by $\Omega(F)$, is
defined as the set of points $x\in X$ for which the following statement
is true: For every neighbourhood $U$ of $x$ and every $T>0$ there is a
$t>T$ such that $F^t(U)\cap U \neq \emptyset$.

A sequence $(x_i)$ of points in $X$ is called an
$\varepsilon-$pseudoorbit for $F$ if for all indices $i$
\[
 d(F(x_{i-1}),x_i) < \varepsilon ,
\]
where $d(\cdot,\cdot)$ denotes the metric on $X$.

The set of chain-recurrent points for $F$, denoted by ${\cal R}(F)$, is
defined as the set of points $x\in X$ for which the following statement
is true: For every $\varepsilon >0$ there is an
$\varepsilon-$pseudoorbit which starts and ends in $x$.

On ${\cal R}(F)$ the following equivalence relation, called chain
equivalence $\sim_{{\cal R}(F)}$, is defined: For $x,y\in {\cal R}(F)$
we have $x\sim_{{\cal R}(F)} y$ if for every $\varepsilon >0$ there is
an $\varepsilon-$pseudo\-orbit from $x$ to $y$ and an
$\varepsilon-$pseudoorbit from $y$ to $x$. The equivalence class
containing $x$ is denoted by $[x]_{{\cal R}(F)}$.
\tme{Def}{5-2}

\tma{Pro}{5-3}
Let a normal possibilistic system
$\htPp:X\times X\rightarrow E$ be motivated by a deterministic system
$F:X\rightarrow X$ ($X$ compact, $\htPp$ upper semicontinuous, $F$
continuous). Then we have
\[
  \Omega(F) \subseteq R_{\htPp} \subseteq {\cal R}(F) .
\]
\tme{Pro}{5-3}

\Proo
If $x\in \Omega(F)$ there are by definition a sequence $(y_n)$ of points
in $X$ and a sequence $(t_n)$ of integers such that $y_n\rightarrow x$
and $F^{t_n}(y_n)\rightarrow x$ as $n\rightarrow\infty$. We know that
$\htPp(y_n,F^{t_n}(y_n)) = \htE$, and since $\htPp$ is normal, this
means that $T_{\htPp}(y_n,F^{t_n}(y_n)) = \htE$, too. Upper
semicontinuity leads to $T_{\htPp}(x,x) = \htE$, and therefore $x\in
R_{\htPp}$.

The second inclusion can be shown in the following way: Suppose
$x\not\in {\cal R}(F)$. This means that there is an $\varepsilon > 0$
such that there is no $\varepsilon-$pseudoorbit from $x$ to $x$, or ---
in other words --- that for every sequence $(x_i)_{0\leq i \leq N}$ with
$x_0=x_N=x$, $N$ arbitrary, there is a $j$ with $0\leq j<N$ such that
$d(F(x_j),x_{j+1})\geq \varepsilon$.
But, because of upper semicontinuity of $\htPp$, this can be reformulated to
the statement that there is a $c < \htE$ such that for every sequence
$(x_i)_{0\leq i \leq N}$ with $x_0=x_N=x$, $N$ arbitrary, there is a $j$
with $0\leq j<N$ such that $\htPp(x_j,x_{j+1})\leq c$. This implies
$\htPp_N(x,x)\leq c$ for every $N$ and therefore $T_{\htPp}(x,x)\leq c$,
showing that $x\not\in R_{\htPp}$.
\Prooe

The arguments of the second part of the proof show that
$T_{\htPp}(x,y)=\htE$ implies that for every $\varepsilon > 0$ there is
an $\varepsilon -$pseudoorbit starting in $x$ and ending in $y$. If the
reverse were true, this would mean that $R_{\htPp} = {\cal R}(F)$ and
even $[x]_{\htPp}=[x]_{{\cal R}(F)}$ for every $x\in {\cal R}(F)$.
However, this equality does not hold in all situations, but in some
important cases, as the next Propositions will show (see also \cite{Ki2}).

\tma{Pro}{5-4}
Let a normal possibilistic system $\htPp:X\times X\rightarrow E$ be
motivated by a deterministic system $F:X\rightarrow X$ ($X$ compact,
$\htPp$ upper semicontinuous, $F$ continuous).
If for every $\chi>0$ there is an $\varepsilon>0$ such that every
$\varepsilon-$pseudoorbit $(y_t)$ in $[x]_{{\cal R}(F)}$ fulfills
$d(y_t,F^t(y))<\chi$ for all $t$ and some $y\in [x]_{{\cal R}(F)}$ (the
so called shadowing property) then $[x]_{{\cal R}(F)} = [x]_{\htPp}$.
\tme{Pro}{5-4}

\Proo
The conditions of the Proposition imply that for every $\chi>0$ and
every $x_1,x_2\in {\cal R}(F)$ with $x_1\sim_{{\cal R}(F)} x_2$ there
is an orbit under $F$ starting in $y_1$ and leading to $z_1$ with
$d(x_1,y_1)<\chi$ and $d(x_2,z_1)<\chi$, and an orbit starting in $y_2$
and leading to $z_2$ with $d(x_2,y_2)<\chi$ and $d(x_1,z_2)<\chi$.
Obviously, $T_{\htPp}(y_1,z_1)=\htE$ and $T_{\htPp}(y_2,z_2)=\htE$.
But upper semicontinuity then leads to
$T_{\htPp}(x_1,x_2)=T_{\htPp}(x_2,x_1)=\htE$, showing that
$x_1\sim_{\htPp} x_2$.
\Prooe

\tma{Pro}{5-5}
Let a normal possibilistic system $\htPp:X\times X\rightarrow E$ be
motivated by a deterministic system $F:X\rightarrow X$ ($X$ compact,
$\htPp$ upper semicontinuous, $F$ continuous).
For $A\subset X$ define $c(A)=\inf\{\htPp(y,z): F(y),z\in A\}$. Let
$x\in {\cal R}(F)$ and $c^*$ be the supremum of all $c\in E$ with the
following property:
There is a $\rho>0$ and a cover ${\cal U}$ of $[x]_{{\cal R}(F)}$ such
that
\[ \htPr_{U\in{\cal U}} c(U_{\rho}) > c    ,\]
where $U_{\rho}=\{x\in X: d(x,U)<\rho \}$.

If $c^* = \htE$ then $[x]_{{\cal R}(F)} = [x]_{\htPp}$.
\tme{Pro}{5-5}

\Proo
Suppose that $c\in E$ fulfills the above mentioned property with
some $\rho>0$ and some cover ${\cal U}$.
If $x_1\sim_{{\cal R}(F)} x_2$ there is a $\rho-$pseudoorbit
$(y_i)_{1\leq i\leq N}$ from $x_1$ to $x_2$. Set $j_1=1$ and define
$j_k$ for $k>1$ recursively
in the following way: $y_{j_k-1}$ is the last point of
$(y_i)_{1\leq i<N}$ whose
image under $F$ lies in $U_{k-1}$ where $U_k\in {\cal U}$ is the set
which contains $F(y_{j_k})$.

This construction leads to a sequence
$(y_{j_k})_{1\leq k\leq\tilde{k}}$ with $j_{\tilde{k}}=N$
($\tilde{k}\leq N$).
The sequence $(y_{j_k})_{1\leq k<\tilde{k}}$ has at most one member
in every $U\in {\cal U}$, and for all $k$ with $1\leq k<\tilde{k}$
the points $y_{j_{k+1}}$ and $F(y_{j_k})$ lie in the same set $U_{\rho}$.
But this means according to the definition of $c(U_{\rho})$ that
\[
  T_{\htPp}(x_1,x_2) \geq \htPp_{(\tilde{k}-1)}(x_1,x_2)
  \geq \htPr_{U\in{\cal U}} c(U_{\rho}) > c .
\]
Taking the supremum over all $c$ we see that $T_{\htPp}(x_1,x_2)
= \htE$. This and the analogous statement with reversed roles for
$x_1$ and $x_2$ leads to $x_1\sim_{\htPp} x_2$.
\Prooe

\tma{Cor}{5-6}
Let $a\htm b = \inf \{a,b\}$, and $\htPp$ be continuous. Then
$[x]_{{\cal R}(F)} = [x]_{\htPp}$ for all $x\in {\cal R}(F)$.
\tme{Cor}{5-6}

\Proo
Since $\htPp$ is normal, deterministically motivated, and continuous,
for all $c<\htE$ there is a $\delta>0$ such that $\htPp(y,z)>c$ for all
$y,z\in X$ with $d(F(y),z)<\delta$.

Set $\delta= 3\rho$ and let ${\cal U}$ be a cover of $[x]_{{\cal R}(F)}$
with sets of diameters smaller than $\rho$. Then for all $y,z\in
U_{\rho}$ we have $d(y,z)<3\rho=\delta$ and therefore $c(U_{\rho})>c$.
This is true for any $U\in {\cal U}$, and so
\[
 \htPr_{U\in {\cal U}} c(U_{\rho}) = \inf_{U\in {\cal U}} c(U_{\rho}) > c .
\]
As this construction works for all $c<\htE$ we have $c^*=\htE$.
\Prooe

\tma{Cor}{5-7}
Let $(E,\hta,\htN,\htm,\htE) = (\Eset^-,\max,-\infty,+,0)$ (see Example
\rse{2-7} \ren{2-7.4}). Let the normal, deterministically motivated,
continuous possibility density $\htPp$ be of order $r$, i.e. there is a
$C\in\Eset^-$, $C\neq 0$, and a $\delta^{*}>0$ such that
\[
  \htPp(x,y) \geq C [d(y,F(x))]^r
\]
for all $x,y\in X$ with $d(y,F(x))<\delta^{*}$.

If the Hausdorff dimension of $[x]_{{\cal R}(F)}$ is less than $r$, then
$[x]_{{\cal R}(F)}=[x]_{\htPp}$.
\tme{Cor}{5-7}

\Proo
Recall that the fact that $r$ is larger than the Hausdorff dimension of
$[x]_{{\cal R}(F)}$ implies that the Hausdorff measure of $[x]_{{\cal R}(F)}$
in dimension $r$ is $0$. This means that for every $\varepsilon>0$ and
for every $\delta>0$ there is a cover ${\cal U}$ of $[x]_{{\cal R}(F)}$ such
that $\mbox{diam}(U) < \delta$ for all $U\in{\cal U}$ and
\[
  \sum_{U\in{\cal U}} [\mbox{diam}(U)]^{r} < \varepsilon .
\]

Since $\htPp$ is of order $r$ one can choose $\delta=\frac{1}{3}\delta^{*}$
and thereby guarantee that if $F(y),z\in U_{\delta}$ then
\[
  \htPp(x,y) \geq C [d(y,F(x))]^r > C [\mbox{diam}(U)]^{r} ,
\]
and therefore $c(U_{\delta})>C[\mbox{diam}(U)]^{r}$.

Thus, for every $\varepsilon>0$ and with the choice $\rho=\delta$
we found a cover ${\cal U}$ of $[x]_{{\cal R}(F)}$ such that
\[
  \sum_{U\in{\cal U}} c(U_{\rho})> C\varepsilon .
\]
But this is exactly the condition of Proposition \rse{5-5}, formulated for
$(E,\hta,\htN,\htm,\htE)$ $= (\Eset^-,\max,-\infty,+,0)$.
\Prooe

We now look at stability in the context of possibilistic systems.

\tma{Def}{5-8}
For a possibility density $\htSs(x)$ with values in an absorptive
semiring, the set
\[
  {\cal I}_{\htSs} = \{x\in X:\htSs(x) = \htN \}
\]
is called the set of impossible points, and the set
\[
  {\cal T}_{\htSs} = \{x\in X:\htSs(x) = \htE \}
\]
is called the set of totally possible points.
\tme{Def}{5-8}

Assume that in a possibilistic system at some time a certain set of
states is regarded as totally possible whereas all other states are
impossible. If for all future times the set of totally possible states
does not change then this set has a stability property which we call
possibilistic stability:

\tma{Def}{5-9}
Let a possibilistic system be given by a transition possibility density
$\htPp(x,y)$ with values in an absorptive semiring. A set $A\subset X$
is called possibilistic stable if for all $t\in\Nset$
\[
  {\cal T}_{O^*_{\htPp_t}\chi_{A}} = A .
\]
\tme{Def}{5-9}

\tma{Rem}{5-10}
\rm
Definition \rse{5-9} leads to a subdivision of equivalence classes
$[x]_{\htPp}$ into stable and non-stable classes of normal basis points.
It is easy to see that a stable equivalence class is characterized by
the following property: There is no $y\in [x]_{\htPp}$ and no $z\not\in
[x]_{\htPp}$ such that $T_{\htPp}(y,z) = \htE$. Still another
characterisation of stable classes is the following: $[x]_{\htPp}$ is
stable iff ${\cal T}_{\Phi_{\htPp,x}} = [x]_{\htPp}$.

One can show that every possibilistic stable set includes at least
one stable class of normal basis points.
\tme{Rem}{5-10}

\tma{Rem}{5-11}
\rm
For the equivalence classes $[x]_{{\cal R}(F)}$ of chain recurrent
points of a deterministic system $F$, stability can be defined in the
following way: $[x]_{{\cal R}(F)}$ is called stable if for all $z\not\in
[x]_{{\cal R}(F)}$ there is an $\varepsilon>0$ such that no
$\varepsilon$-pseudoorbit leads from $[x]_{{\cal R}(F)}$ to $z$.
Ruelle \cite{Ru2} calls a stable $[x]_{{\cal R}(F)}$ an attractor of the
deterministic system $F$.

In cases in which for a deterministically motivated possibilistic
system the classes $[x]_{{\cal R}(F)}$ and $[x]_{\htPp}$ coincide, the
two definitions of stability correspond.
\tme{Rem}{5-11}

We sum up the findings of this section: In a deterministically motivated
normal possibilistic system the long term behaviour is governed by the
normal basis points, which include the non-wandering points of the
deterministic system and are contained in the set of chain-recurrent
points. The stability of classes of normal basis points can be
recognised by the shape of the maxima in the corresponding
eigenfunctions of the pseudointegral operator which has the transition
possibility as a kernel. These eigenfunctions can be calculated from the
transitive closure of the transition possibility. The definition of
stable classes of basis points is similar --- and in certain cases
equivalent --- to Ruelle's attractor definition. This shows that this
attractor definition is robust to uncertainties which can be modelled
possibilistically.

\rsec{Connections between possibilistic \protect\\ and
      stochastic systems}{6}

Certain possibilistic systems can be used to obtain information about
stochastic systems. As a first example we discuss the application of
systems with transition possibility densities in the semiring
$(E,\hta,\htN,\htm,\htE)$=$([0,1],\max,0,\min,1)$ (see Example \rse{2-7}
\ren{2-7.3}) to stochastic systems in which the transition probabilities
have compact support.

More specifically, let $(P^{(\alpha)}_x)$ be a family of transition
probabilities on a compact metric state space $X$, depending
continuously --- in the topology of weak convergence --- on $x$ and on
the parameter $\alpha\in [0,1]$ such that
\[
   \mbox{supp }P^{(\alpha)}_x \subset \mbox{ supp }P^{(\beta)}_x\ 
   \mbox{ strictly } \quad \mbox{ if } \alpha > \beta
\]
for all $x\in X$.

Under the present conditions there is at least one family
$(S^{(\alpha)*})$ of invariant measures of the stochastic systems
$P^{(\alpha)}_x$ with a continuous dependence on $\alpha$.

Now define for all $x,y\in X$
\[
   \htPp(x,y) = \max \{\alpha: y\in \mbox{ supp }P^{(\alpha)}_x \} .
\]
Interpret these as transition possibility densities in the semiring
$(E,\hta,\htN,\htm,\htE)$= \linebreak $([0,1],\max,0,\min,1)$.

Further define for all $x\in X$
\[
   \htSs(x) = \max \{\alpha: x\in \mbox{ supp }S^{(\alpha)*} \} .
\]

\tma{Pro}{6-1}
With the notation introduced above, the function $\htSs$ is an invariant
possibility density of the possibilistic system defined by $\htPp$.
\tme{Pro}{6-1}

\Proo
We start with the equation of invariance
\[
   S^{(\alpha)*}(A) = \int_X P^{(\alpha)}_y(A)\ S^{(\alpha)*}(dy) ,
\]
$A\in{\cal A}$.

So we can write
\[
   \htSs(x) = \max \{\alpha:
      x\in \mbox{ supp }\int_X P^{(\alpha)}_y(.)\ S^{(\alpha)*}(dy) \} .
\]
But $x\in \mbox{ supp }\int_X P^{(\alpha)}_y(.)\ S^{(\alpha)*}(dy) $
iff there is at least one $y\in X$ such that
\[
x\in \mbox{ supp }P^{(\alpha)}_y \mbox{\ \ and\ \ }
y\in \mbox{ supp }S^{(\alpha)*}.
\]
This means that $\alpha$ must be smaller or equal to the minimum of
$\htPp(y,x)$ and $\htSs(y)$ for at least one $y$, and hence
\begin{eqnarray*}
   \htSs(x) & = & \max_{y\in X}\ \min \{ \htSs(y), \htPp(y,x) \} \\
            & = & \htIn_X \htSs(y) \htm \htPp(y,x) \htm dy \\
            & = & (O^*_{\htPp}\htSs)(x)  .
\end{eqnarray*}
\Prooe

\tma{Exa}{6-2}
\rm
An important special case of the situation described above is the case
where for all $x\in X$
\[
   P^{(1)}_x = \delta_{F(x)}
\]
with a continuous function $F:X\rightarrow X$. The stochastic system can
then be regarded as a stochastically perturbed deterministic system
given by $F$. The transition $\alpha\rightarrow 1$ describes vanishing
noise strength.

A concrete example would be a dynamical system with additive noise,
\[
   x_{t+1} = F(x_t) + (1-\alpha) \xi_t ,
\]
where $\xi_t$ are uncorrelated random variables distributed according to
a probability measure with compact support.

The possibilistic system $\htPp$ is obviously deterministically
motivated by $F$.

Putting together the results of Section \rse{4} and Corollary \rse{5-6}
we see from Proposition \rse{6-1} that the limiting invariant measures
$S^{(1)*}$ are concentrated on classes of chain recurrent points --- in
agreement with other, more detailed results \cite{Ru2,Ki2}.

The support of the invariant measure in the case of non-vanishing noise
strength $\varepsilon = 1 - \alpha$ can be obtained from the corresponding
eigenfunction $\htSs$ of $O^*_{\htPp}$ as the level cut
\[
  \{ x\in X: \htSs(x) \geq 1-\varepsilon \} .
\]
\tme{Exa}{6-2}

A second way of connecting possibilistic systems to stochastic systems
makes use of exponential estimates which belong to the so called
``large deviation'' method \cite{Va1,DZ1}. The famous large deviation
principle can 
conveniently be expressed in the language of possibility measures:

\tma{Def}{6-3}
A family of probability measures $(P^{(\varepsilon)})_{\varepsilon>0}$
on a complete separable metric space $X$
is said to obey the large deviation principle with possibility measure $\htP$
with values on the semiring $(E,\hta,\htN,\htm,\htE)$=
$(\Eset^{-},\max,-\infty,+,0)$ (see Example \rse{2-7}
\ren{2-7.4}) if $\htP$ has an upper semicontinuous
possibility density $\htPp$ and if for all subsets $A\subset X$
\[
  \htP(A^{o}) \leq
  \lim_{\varepsilon\rightarrow 0}\inf \varepsilon \log
                                            P^{(\varepsilon)}(A) \leq
  \lim_{\varepsilon\rightarrow 0}\sup \varepsilon \log
                                            P^{(\varepsilon)}(A) \leq
  \htP(\overline{A})
\]
where $A^{o}$ is the interior of $A$ and $\overline{A}$ its closure.
\tme{Def}{6-3}

Note that the fact that probability measures are normalised implies that
the possibility measures appearing in the large deviation property must
be normal.
The negative of $\htPp$ is usually called rate function. A frequently imposed
further condition is that the level cuts of $\htPp$ be compact. Since we are
assuming a compact $X$ anyway, this condition is fulfilled automatically here.

If the rate function is continuous then we have because of compact $X$:
\[
  \lim_{\varepsilon\rightarrow 0} \varepsilon \log
                                        P^{(\varepsilon)}(A) =
  \htP(A)
\]
for all open $A\subset X$.

Many of the results obtained by large deviation methods have a
suggestive formulation in possibilistic language. Here is an
example:

\tma{Pro}{6-4}
{\sf (Varadhan's generalised Laplace method)}
Let $(P^{(\varepsilon)})$ have the large deviation property with
possibility measure $\htP$. Let $G:X\rightarrow \Eset^-$ be a
continuous function. Then
\[
  \lim_{\varepsilon\rightarrow 0} \varepsilon \log
      \int_X \exp \left( \frac{G(x)}{\varepsilon} \right)
                  P^{(\varepsilon)}(dx) =
  \htIn_X G(x)\htm\htP(dx) .
\]
\tme{Pro}{6-4}

We now look at a family of stochastic dynamical systems whose transition
probabilities have a large deviation property:

\tma{Pro}{6-5}
Assume that a family $(P_x^{(\varepsilon)})$ of transition probabilities
on a compact state space $X$ with continuous dependence on $x\in X$ has
the large deviation property with transition possibility $\htP_x$
and that the possibility density is continuous. If there is a family
$(S^{(\varepsilon)*})$ of invariant measures of the stochastic systems
$P_x^{(\varepsilon)}$ that has the large deviation property with a
possibility measure $\htS^*$ then $\htS^*$ is an invariant possibility
of the possibilistic system defined by $\htP_x$, i.e.,
\[  \htS^*(A) = \htIn_X \, \htP_y(A) \htm \htS^*(dy) \]
for all open $A\in{\cal A}$.
\tme{Pro}{6-5}

\tma{Rem}{6-6}
\rm
This proposition is a simple consequence of the Laplace-Vara\-dhan type
approximation of integrals. However, it does by no means prove the fact
that invariant measures for the stochastic systems actually have the
large deviation property. Proofs of this property or similar estimates
exist for a wide range of systems but require considerably more work
\cite{Ki1,Ha1}. They go back to the work of Wentzell and Freidlin
\cite{FW1} on random
perturbations of dynamical systems, which is the continuous time
analogue of what we are discussing here. From this background
the negative of the possibility density $\htSs$ of $\htS^*$ has the name
quasipotential.

In the case where the stochastic systems are random perturbations of a
deterministic system $F$ the corresponding possibilistic system is of
course deterministically motivated by $F$.
\tme{Rem}{6-6}

Quasipotentials have been discussed in the physical literature as
nonequilibrium generalisations of thermodynamic potentials (see
\cite{Gr1} for a review).
The discrete time version \cite{Ki1,HG1,RT1} has been used
successfully to investigate the 
influence of noise on renormalisation schemes in the context of
transitions from regular to chaotic behaviour \cite{HG2}, and other universal
aspects of the influence of noise on bifurcations \cite{HTG}.

In the physical literature one usually finds heuristic derivations of
quasipotentials instead of a strict application of the mathematical
results from the Wentzell-Freidlin tradition. These derivations use
approximations of the La\-pla\-ce type, not unlike our motivation for
Proposition \rse{6-5}. In this way, for the case of Gaussian noise, the
following equation was found in \cite{RT1} for determining the quasipotential
$(-\htSs)$:
\[ (-\htSs(y)) = \min_{x} \left[ (-\htSs(x)) + \frac{1}{2} (y-F(x))^2
                          \right]          .  \]
In our language of pseudomeasures this is the statement that $\htSs$ is
an eigenfunction of the pseudolinear operator $O^*_{\htPp}$,
$\htPp(x,y) = -\frac{1}{2} (y-F(x))^2$, with eigenvalue $\htE=0$ ---
in agreement with Proposition \rse{6-5}.

Applying the results of Section \rse{4} and Section \rse{5}, one sees
immediately several facts about quasipotentials which were so far only
accessible by following the details of the proofs of the
Wentzell-Freidlin approach, and not by the heuristic approaches. This
refers in particular to the definition of basis classes, their
importance for the construction of quasipotentials, and their relation
to sets defined by properties of deterministic systems (like
non-wandering sets, chain-recurrent sets, and attractors).

The connection between quasipotentials and the possibilistic algebra
brings another advantage, namely a systematic approach to the numerical
computation of quasipotentials. In such computations the eigenfunction
problem is transformed into an $N$-dimensional eigenvalue problem based
on the semi\-ring $(\Eset^{-},\max,-\infty,+,0)$ by discretising the space
$X$ into a finite space of $N$ points. There are well-known and
well-studied numerical algorithms for this finite dimensional problem
(which is equivalent to the problem of finding shortest paths in a
graph \cite{GM2}).

\tma{Rem}{6-7}
\rm
Consider the eigenproblem for eigenvalue $\htE$ on the semimodule
$(\Eset^{-})^N$ with operations $\hta=\max$ and $\htm=+$ and definite
transition possibilities $\htPp_{ij}$:
\[
  \htSs_j = \htSu_{i=1}^N \htSs_{i} \htm \htPp_{ij}
          = \max_{i=1,\dots,N} (\htSs_i + \htPp_{ij}) .
\]
Interpreting $-\htPp_{ij}$ as the length of the arc $(ij)$ in a directed
graph, the computation of an $\htSs_j$ amounts to finding the shortest
path (whose length then is -$\htSs_j$) between a basis point and $j$.

Following the pseudolinear formalism one can write down a general
algorithm for solving the eigenproblem which is analogous to the
Gauss-Seidel algorithm of linear algebra. The resulting algorithm is
equivalent to Ford's algorithm \cite{Fo1} for solving the shortest
path problem .
The computing time required for that algorithm is at most ${\cal
O}(N^3)$.

If $\htPp$ is not definite, the computation of the eigenfunctions has to
be preceded by the computation of the eigenvalue $\lambda$, which is the
negative of the minimal cycle mean in graph theoretic language (see
Remark \rse{4-24}). An appropriate method is Karp's algorithm (see
\cite{FG1}), again
with a computing time of ${\cal O}(N^3)$.

These combinatorial matrix-type methods compete with certain iterative
methods which take a time of order ${\cal O}(N^2)$ for each step of the
iteration (see \cite{Nu1}).

However, if $\htPp$ is bounded above by $\htE=0$ (i.e., there are no
arcs of negative length in the graph), there is a faster combinatorial
algorithm, called Dijkstra's algorithm \cite{Di1}. Computing times are of order
${\cal O}(N^2)$ in this case, so that the iterative methods mentioned
above are no longer an attractive alternative.

For a deterministically motivated problem it is possible to further
reduce computing times by concentrating on those arcs $(ij)$ for
which the point with label $j$ lies in a small neighbourhood of the
deterministic image of the point with label $i$, and this algorithm
appears to be the most efficient general method for the computation of
quasipotentials.
\tme{Rem}{6-7}

We know from Section \rse{4} that there are pseudolinearly independent
invariant possibility densities for a normal possibilistic system as
soon as there is more than one equivalence class of basis points. For
this case we have to discuss, which pseudolinear combinations of
invariant possibility densities qualify for being quasipotentials.

\tma{Pro}{6-8}
Consider the situation of Proposition \rse{6-5}. If $\htP_x$ has the
possibility density $\htPp(x,y)$ then the invariant possibility density
$\htSs$ characterising the large deviation property of
$(\htS^{(\varepsilon)*})$ can be written as
\[
  \htSs(x) = \htIn_{R^s_{\htPp}} \htSs(a) \htm T_{\htPp}(a,x) \htm da
\]
where $R^s_{\htPp}$ is the union of stable classes.

The values of $\htSs(a)$ for $a\in R^s_{\htPp}$ have to fulfill the
following equations:
\[
  \htIn_{R^s_{\htPp}-[a]_{\htPp}}
        \htSs(a) \htm T_{\htPp}(a,b) \htm db   =
  \htIn_{R^s_{\htPp}-[a]_{\htPp}}
        \htSs(b) \htm T_{\htPp}(b,a) \htm db
\]
for all $a\in R^s_{\htPp}$.
\tme{Pro}{6-8}

\Proo
Since $S^{(\varepsilon)*}$ is an invariant measure for the transition
probability measures $P_x^{(\varepsilon)}$, we have for every $A\in
{\cal A}$:
\[
  \int_A \ \int_{X-A} P_a^{(\varepsilon)}(dy) S^{(\varepsilon)*}(da)
   =
  \int_A \ \int_{X-A} P_y^{(\varepsilon)}(da) S^{(\varepsilon)*}(dy) .
\]
On the level of the corresponding possibility densities this reads:
\[
  \htIn_{X-A} \htSs(a) \htm \htPp(a,y) \htm dy
   =
  \htIn_{X-A} \htSs(y) \htm \htPp(y,a) \htm dy .
\]
In this equation, $\htPp$ can be replaced by $\htPp_N$ for any $N\in
\Nset$, and therefore by $T_{\htPp}$.

We set $A=\left\{x\in X: T_{\htPp}(a,x) \leq
\htIn_{R^s_{\htPp}-[a]_{\htPp}} T_{\htPp}(a,y)\htm dy \right\}$ where
$a\in R^s_{\htPp}$. We see that the domain of pseudointegration on the
left hand side can be replaced by $R^s_{\htPp}-[a]_{\htPp}$.

On the right hand side we insert
\[
  \htSs(y) = \htIn_{R_{\htPp}} \htSs(b) \htm T_{\htPp}(b,y) \htm db ,
\]
which is a consequence of Proposition \rse{4-18}. Since $y\in X-A$, the
pseudointegration can be restricted to $R^s_{\htPp}-[a]_{\htPp}$, and we
have
\[
  \htIn_{R^s_{\htPp}-[a]_{\htPp}} \htSs(a) \htm T_{\htPp}(a,y) \htm dy
   =
  \htIn_{X-A} \htIn_{R^s_{\htPp}-[a]_{\htPp}} \htSs(b) \htm
            T_{\htPp}(b,y) \htm db \htm T_{\htPp}(y,a) \htm dy .
\]
Carrying out the pseudointegral over $y$ we arrive at the stated
relation between the coefficients $\htSs(a)$.
\Prooe

\tma{Rem}{6-9}
\rm
In the case that there is a finite number $L$ of stable basis classes,
Proposition \rse{6-8} gives $L$ pseudolinear equations for the
coefficient for the $L$ independent eigenfunctions. However, only $L-1$
of these equations are pseudolinearly independent. The coefficients are
uniquely determined if we add the normalisation condition $\htIn_X
\htSs(x) = \htE = 0$.

Freidlin and Wentzell describe a graph method for solving the
condition from Proposition \rse{6-8}. This is quite natural since the
condition is similar to a Kirchhoff rule, for which graph methods have a
long tradition.
\tme{Rem}{6-9}

In the literature about quasipotentials (e.g. \cite{GT1,HG1}) a
connection to Hamiltonian systems has been established and
exploited. In the same spirit we 
formulate a connection between symplectic maps (see e.g. \cite{Me1})
and possibilistic systems 
evaluated on the semiring $(\Eset^{-}$, $\max,-\infty,+,0)$. We briefly
sketch some aspects of this connection; details can be found in the
literature mentioned above.

Let $G:X\times X \rightarrow \Rset$, $X\subset \Rset^d$, be a
differentiable function such that the map $q'\mapsto \partial_1 G(q,q')$
is a diffeomorphism for all $q\in X$. Let $G$ be a generating function
for a symplectic map $\tilde{H}:X\times\Rset^d \rightarrow
X\times\Rset^d$, $(q,p)\mapsto (q',p')$, i.e.
\begin{eqnarray*}
 p' & = & \phantom{-} \partial_2 G(q,q') \\
 p \phantom{'}  & = & - \partial_1 G(q,q')
\end{eqnarray*}
where $\partial_i$ denotes differentiation with respect to the $i$th
argument.
Let $\Pi:X\times \Rset^{d} \rightarrow X$ be the projection $\Pi(q,p)=q$.

\tma{Pro}{6-10}
Define a possibilistic system on $X$ by a possibility density with
values in $(\Eset^{-}$, $\max,-\infty,+,0)$:
\[
  \htPp = - G
\]
where $G$ is a generating function for the symplectic map $H$.

If $(q_0,\dots,q_N)$ is a sequence of points in $X$ for which
\[
  \htPp_N(q_0,q_N) = \htPr_{i=1}^N  \htPp(q_{i-1},q_i)
                   = \sum_{i=1}^N  \htPp(q_{i-1},q_i)
\]
then it is the projection of the trajectory under $H$ which starts in
the point $v_0=(q_0,-\partial_1 G(q_0,q_1))$: $q_i=\Pi(H^i(v_0))$.
\tme{Pro}{6-10}
This follows directly from making use of differentiability in the
maximisation required for the determination of $\htPp_N$.

The determination of $T_{\htPp}(x,y)$ requires then a maximisation among
all trajectories under $H$ which start with a $q$-component $x$ and end
with a $q$-component $y$.

The normal basis points of a definite possibilistic system are
those recurrent points of the symplectic map along whose trajectories
the sums of $G(q,q')$ have the globally minimal value $0$.

If $\htPp$ is normal then the space $\{(q,p):p=0\}$ is invariant under
$H$. This follows from the fact that because of normality there is for
each point $q$ a point $q'$ such that $G(q,q')$ attains its minimum $0$,
which implies $\partial_2 G(q,q') = 0$ and $\partial_1 G(q,q') = 0$.
Note that the invertibility condition on $\partial_1 G$ can only
be fulfilled if the normal possibilistic system is deterministically
motivated by a map $F:X\rightarrow X$. On $\{(q,p):p=0\}$ the symplectic
map reduces to $H(q,0)=(F(q),0)$. The recurrent points corresponding to
basis points have $p=0$, too.

Using the connection between $p$-values and derivatives of $\htPp$ one
can write
\[
  \partial_2 \htPp_N(q_0,q_N) =  p_N
\]
where $H^N(v_0)=(q_N,p_N)$. Analogously,
\[
  \partial_1 \htPp_N(q_0,q_N) = p_0
\]
These equations can be used to determine eigenfunctions $T_{\htPp}(a,x)$
and $T_{\htPp}(x,a)$, $a$ a basis point, by integrating the
$p$-values along invariant manifolds emanating from the point $(a,0)$.
Without going into the details of Lagrangian manifolds (see
e.g. \cite{Ar1}) we mention that
generically these invariant manifolds have tangling bends which lead to
accumulating Maxwell points at which the eigenfunctions are not
differentiable (see \cite{HG1}; many interesting aspects of this phenomenon
have been studied in the case of quasipotentials for continuous time systems
\cite{GT1,Ja1,MS2,DM1}).

The relation between symplectic maps and possibilistic systems
suggests a potentially wide field of applications where idempotent
pseudolinear algebra can be used in variational problems and
Hamiltonian mechanics.
 
An early example for such an application can be found
in the method of effective potentials for one-dimensional infinite chains of
atoms (such as the Frenkel-Kontorova model), introduced by Chou and Griffiths
\cite{CG1}. This method can be translated into the language of uncertain
dynamical systems: We ask for the possibility that the position of
an atom at site $t+1$ in the chain is $x_{t+1}$, knowing that the position of
the atom at site $t$ is $x_{t}$. This possibility is assessed at a value
$-K(x_{t},x_{t+1})$ in the semiring $(\Eset^{-},\max,-\infty,+,0)$ where
$K(x_{t},x_{t+1})$ is the energy which is added to the system when adding the
atom number $t+1$ at position $x_{t+1}$ to a semi-infinite chain which ends
with the $t-$th atom at position $x_{t}$.

In general, the possibilistic system
defined in this way is not normal and not even definite, and therefore it does
not have an invariant possibility density. But the semiring has
an invertible pseudomultiplication and is radicable and absorptive, and if $K$
is bounded, Proposition \rse{4-23} and Remark \rse{4-24} are applicable.
Thus, if $\lambda$ is the minimal cycle mean $\ds \frac{1}{N}
\sum_{i=1}^{N}K(q_{i-1},q_{i})$ over cyclic sequences $(q_{1},\dots,q_{N})$
then $\tilde{k}(x,y)=\lambda - K(x,y)$ is definite, and $O^{*}_{\tilde{k}}$
has the eigenvalue $\htE=0$ whose eigenfunctions can be called 
conditionally invariant possibility densities in analogy to the notion of
conditionally invariant measures of deterministic systems \cite{PY1,Te1}. From
Section \rse{4} we know that the basis classes (Definitions \rse{4-7} and
\rse{4-10}) for $\tilde{k}$ play a crucial
role in finding the conditionally invariant possibility densities.

Interestingly, all the concepts introduced in order to describe the
possibilistic system have useful interpretations in the original context of
the underlying system of atoms: The eigenvalue $\lambda$ is the minimal energy
per atom in the chain. The basis classes $[a]_{\tilde{k}}$, $a\in
R_{\tilde{k}}$, are the pure ground states of the chain. The conditionally
invariant possibility density $\Phi_{\tilde{k},a}(x)=T_{\tilde{k}}(a,x)$,
$a\in R_{\tilde{k}}$, is an effective potential acting on the right-most atom
with position $x$ in a semi-infinite chain extending to the left,
asymptotically approaching the ground state configuration $[a]_{\tilde{k}}$.
Similarly, $\Psi_{\tilde{k},a}(x)=T_{\tilde{k}}(x,a)$ is an effective
potential for the left-most atom in a semi-infinite chain extending to the
right. The two effective potentials $\Phi_{\tilde{k},a}(x)$ and
$\Psi_{\tilde{k},a}(x)$ can be used to compute excitation energies of defects,
see \cite{CG1}.

This last example shows that the concept of possibilistic dynamical systems
can be useful not only as a method for dealing with uncertainties 
but beyond our initial motivation also as an approach to general variational
problems.  

\newpage

\section*{Acknowledgement}
I am grateful to Robert MacKay for stimulating discussions.

Work on this paper started during a stay at the Mathematics Institute of
the University of Warwick on a grant by the European Commission under their
Human Capital and Mobility scheme. Financial support by the {\em Deutsche
Forschungsgemeinschaft} (SFB 237) is also gratefully acknowledged.

\end{document}